\documentclass[10pt]{article}
\usepackage{graphicx}

\begin{document}
 \title{Different effect of the hyperons $\Lambda$ and $\Xi$ \\ on the
nuclear core  }
\author{Yu-Hong Tan$^1$,Ping-Zhi Ning$^2$\\
1. Theoretical Physics Division,
          Nankai Institute of Mathematics, \\ Nankai University,
   Tianjin 300071, P.R.China \\
  2. Department of Physics, Nankai University,
   Tianjin 300071, P.R.China }
\date{}
 \maketitle
\begin{abstract}
We demonstrate the different effect of strange impurities
($\Lambda$ and $\Xi$) on the static properties of nuclei within
the framework of the relativistic mean-field model. Systematic
calculations show that the gluelike role of $\Lambda$ hyperon is
universal for all $\Lambda$-hypernuclei considered. However,
$\Xi^-$ hyperon has the gluelike role only for the protons
distribution in nuclei, and for the neutrons distribution $\Xi^-$
hyperon plays a repulsive role. On the other hand, $\Xi^0$ hyperon
attracts surrounding neutrons and reveals a repulsive force to the
protons. Possible explanations of the above observation are
discussed.

\noindent{\bf PACS numbers} . 21.80.+a - Hypernuclei. \\{\bf PACS
numbers}. 24.10.Cn - Many-body theory.
\end{abstract}

\section{Introduction}
 Change of bulk properties of nuclei under the presence of strange
 impurities, like the lambda hyperon ($\Lambda$), is an
 interesting subject in hypernuclear physics. Since a $\Lambda$
 does not suffer from Pauli blocking, it can locate at the center
 of a nucleus, then the $\Lambda$ attracts surrounding nucleons
 (the gluelike role of $\Lambda$) and makes the nucleus
 shrink. One might expect only a bit change of the size for most
 of the nuclei. However, significant shrinkage of hypernuclei size
 could be expected when a $\Lambda$ is added to loosely bound
 light nuclei such as $^6{\rm Li}$$^{\cite{1983,2,3}}$. Recently,
 the experiment KEK-PS E419 has found clear evidence for this shrinkage of
  hypernucleus $^7_\Lambda{\rm Li}$$^{\cite{npa2001, prc2001}}$.

In order to obtain a more profound understanding of the gluelike
role of strange impurities in nuclei, it is necessary to consider
other strange impurities, like the sigma ($\Sigma^-$) and cascade
($\Xi^-, \Xi^0 $). The behavior of these hyperons in the nuclear
medium as well as the hyperon-nucleus potential, is of particular
importance for this study. However, the $\Sigma$-nucleus potential
has been still unclear until more recently because experimental
information is limited$^{\cite{6019}}$. For the sake of the
improvement of this situation, a new experiment at KEK is carried
out to measure the inclusive ($\pi^-, {\rm K}^-$) spectrum, which
is sensitive to the $\Sigma$-nucleus potential$^{\cite{new7}}$.
The result shows that a strongly repulsive $\Sigma$-nucleus
potential is required to reproduce the observed spectrum. So, we
have reason to believe that $\Sigma$ hyperon does not have any
gluelike role and can not make the nucleus shrink. Next in mass
are the $\Xi^-$ and $\Xi^0$ hyperons. Experimental evidence
suggests that $\Xi$-nucleus potential is attractive
$^{\cite{6025}}$. Therefor we may only consider $\Lambda$- and
$\Xi$-hypernuclei in this work. Our purposes are: (i)to test the
universality of the gluelike role of $\Lambda$ impurity in a
variety of $\Lambda$-hypernuclei which may not be loosely bound
light nuclei; (ii) to see whether or not the $\Lambda$- and
$\Xi$-impurities behave the same, in view of the fact that both
$\Lambda$- and $\Xi$-nuclear potentials are attractive; (iii) to
predict the properties of the $\Xi$-impurities in
$\Xi$-hypernuclei.

To accomplish these, a standard approach to the subject is the
relativistic mean-field (RMF) model, to which a brief description
for the hypernuclei is given in Sec.II. In Sec.III, after testing
the validity of force parameters used in the RMF model, systematic
calculations are performed for $\Lambda$-hypernuclei and the
universality of the gluelike role of $\Lambda$ impurity is
revealed. In Sec.IV, we provide the RMF results for
$\Xi$-hypernuclei and different effects of $\Xi^-$ and $\Xi^0$ on
the nucleus are discussed. A brief summary and conclusions are
drawn in Sec.V.

\section{The RMF Model}

The relativistic mean-field model (RMF) has been used to
 describe nuclear matter, finite nuclei, and hypernuclei successfully.
Here, we start from a Lagrangian density of the form
\begin{equation}
  {\cal L}={\cal L}_N+{\cal L}_Y,
\end{equation}
where, $Y=\Lambda$, or $\Xi^-$, or $\Xi^0$, ${\cal L}_N$ is the
standard Lagrangian of the RMF model
\begin{eqnarray}
{\cal L}_N&=&\bar{\psi}_N(i\gamma_\mu\partial^\mu-M_N-g_{\sigma N
}\sigma-g_{\omega N}\gamma_\mu\omega^\mu \nonumber \\
&-&\frac{1}{2}g_{\rho N}\gamma_\mu \vec{\tau}_N\cdot{\vec
{\rho}~^\mu} -
e\gamma_\mu \frac{1+\tau_{3,N}}{2} A^\mu)\psi_N \nonumber \\
&+&\frac{1}{2}(\partial_\mu\sigma\partial^\mu \sigma
-m^2_\sigma \sigma^2)\nonumber \\
&-&\frac{1}{3}b\sigma^3-\frac{1}{4}c\sigma^4\nonumber \\
&-&\frac{1}{4}\Omega_{\mu\nu}\Omega^{\mu\nu}+\frac{1}{2}m^2_\omega\omega_\mu\omega^\mu
\nonumber \\
&-&\frac{1}{4}{\vec{ R}}_{\mu\nu}\cdot{\vec
{R}}^{\mu\nu}+\frac{1}{2}m_\rho^2{\vec {\rho}}_\mu\cdot{\vec
{\rho}}~^\mu\nonumber \\
&-&\frac{1}{4}H_{\mu\nu}\cdot H^{\mu\nu}
\end{eqnarray}
where
\begin{eqnarray}
   \Omega_{\mu\nu}=\partial_\nu\omega_\mu-\partial_\mu\omega_\nu, \nonumber \\
   {\vec {R}}_{\mu\nu}=\partial_\nu \vec{\rho}_\mu-\partial_\mu{\vec{\rho}}_\nu,   \nonumber \\
H_{\mu\nu}=\partial_\nu A_\mu-\partial_\mu A_\nu,
\end{eqnarray}
 It involves nucleons ($\psi_N$), scalar $\sigma$ mesons
($\sigma$), vector $\omega$ mesons ($\omega_\mu$), vector
isovector $\rho$ mesons ($\vec{\rho}_\mu$), and the photon
($A_\mu$). The scalar self-interaction
$-\frac{1}{3}b\sigma^3-\frac{1}{4}c\sigma^4$ are included, as
well. The parametrization of the nucleonic sector (NL-SH)are
adopted from Ref.\cite{18}, the properties of finite nuclei can be
well described.

The Lagrangian density ${\cal L}_\Lambda$ describes the hyperon
$\psi_\Lambda$ and its couplings to mesonic fields includes the
$\omega$-$\Lambda$ tensor coupling term
\begin{eqnarray}
 {\cal L}_\Lambda&=&\bar\psi_{\Lambda}(i \gamma^ \mu \partial_\mu
-m_{\Lambda}- g_{\sigma \Lambda}\sigma -
 g_{\omega\Lambda}\gamma_\mu\omega^\mu)\psi_{\Lambda} \nonumber \\ &+& \frac{f_{\omega \Lambda}}{2m_Y} \bar
 \psi _\Lambda \sigma_{\mu\nu}\cdot\partial^\nu \omega^\mu
 \psi_\Lambda.
\end{eqnarray}
Since $\Lambda$ is neutral and isoscalar baryon, it does'not
couple with the $\rho$ mesons and the photon. We adopt the
parametrization of $\Lambda$ sector from Ref.\cite{ma}:
$g_{\sigma\Lambda}/g_{\sigma N}=0.49, g_{\omega \Lambda
}/g_{\omega N}=0.512, f_{\omega \Lambda}/g_{\omega
\Lambda}=-0.616$. Using these coupling constants, the properties
of $\Lambda$-hypernuclei can be well described$^{\cite{ma,tana}}$.

 The Lagrangian density ${\cal L}_\Xi$ describes
the hyperon $\psi_\Xi$ and its couplings to the $\sigma$, $\omega$
, $\rho$ mesonic fields and the photon field
\begin{eqnarray}
 {\cal L}_\Xi&=&\bar\psi_{\Xi}(i \gamma^ \mu \partial_\mu
-m_{\Xi}- g_{\sigma \Xi}\sigma -
 g_{\omega\Xi}\gamma_\mu\omega^\mu\nonumber \\
 &-&\frac{1}{2}g_{\rho \Xi}\gamma_\mu {\vec {\tau}_\Xi}\cdot{\vec{ \rho}}~^\mu -
e\gamma_\mu \frac{\tau_{3,\Xi}-1}{2} A^\mu)\psi_{\Xi}.
\end{eqnarray}
We fix the coupling constants of $\Xi$, say the one to the vector
fields with the quark model (SU(6) symmetry),
\begin{eqnarray}
 g_{\omega
  \Xi}= \frac{1}{3}g_{\omega N},\\
g_{\rho \Xi}=  g_{\rho N},
\end{eqnarray}
and those to the scalar field with the experimental
information--the optical potential. It turns out that two coupling
constants of $\Xi$, $g_{\sigma \Xi}$ and $g_{\omega \Xi}$, are
strongly correlated because they are fixed by the depth of the
$\Xi$ potential
\begin{equation}
  V^\Xi_0=g_{\sigma \Xi}\sigma^{eq}+g_{\omega
  \Xi}\omega^{eq}
\end{equation}
in the saturation nuclear matter$^{\cite{sun5,9622}}$.
 But the experimental data on
$\Xi^-$-hypernuclei are very little.
 Dover and Gal$^{\cite{dg}}$ analyzing old emulsion data on
 $\Xi^-$ hypernuclei, conclude a nuclear potential well depth of
 $V^\Xi_0=-21\sim-24$ MeV. Fukuda et al$^{\cite{6024}}$ fit to
 the very low energy part of $\Xi^-$ hypernuclear spectrum in the
 ${\rm ^{12}C(K^-,K^+)X}$ reaction in
experiments E224 at KEK, estimate the value of $V^\Xi_0$ between
$-16$ and $-20$ MeV. Recently, E885 at the AGS $^{\cite{6025}}$
indicates a potential depth of $V^\Xi_0=-14$ MeV or less. So the
depth $V^\Xi_0$ of $\Xi$ in nuclear matter is not well fixed on.

\section{The effect of the $\Lambda$ impurity}

We start from calculation of the single-particle energies
 for $\Lambda$ in $\Lambda$-hypernuclei within the framework
 of the RMF model with force parameters taken from Ref.{\cite{ma}}, and
 present the results in Fig.1. It can be seen that the results
are in good agreement with the experimental
data$^{\cite{7,1999,pile}}$. Very small spin-orbit splitting for
$\Lambda$-hypernuclei are also observed. This shows the RMF theory
with the parameter set used for the $\Lambda$-hyperonic sector is
reliable for studying the effect of the $\Lambda$ impurity, and
has a predicting ability.

In order to observe the universality of the gluelike role of the
$\Lambda$ hyperon impurity, an unified RMF calculation is needed
and careful tests should be done. Hence in our calculations
typical hypernuclei between $^7_\Lambda{\rm Li}$ and
$^{209}_\Lambda{\rm Pb}$ are selected. Our results are shown in
table \ref{t1}, in which some results for medium and heavy
hypernuclei have been given in our previous work$^{\cite{tana}}$.
In the table, $-E/A$ (in MeV) is the binding energies per baryon,
$r_{ch}$ is the r.m.s. charge radius, and $r_y$, $r_n$ and $r_p$
are the calculated r.m.s. radii (in fm) of the hyperon ($\Lambda$
or $\Xi$), neutron and proton, respectively. Here, hyperon is at
its $1s_{1/2}$ configuration. The definition of these quantities
can be found in ref.\cite{new19}. For comparison, the results for
normal nuclei are also given. From table \ref{t1}, it can be seen
that for lighter $\Lambda$-hypernucleus, the size of the core
nucleus in a hypernucleus is smaller than the core nucleus in free
space (i.e., normal nucleus). Although there are only a bit change
in the core nucleus due to the presence of $\Lambda$ impurity. For
instance, the r.m.s. radius $r_n$ ($r_p$) of the neutrons
(protons) decreases from 2.32 fm (2.37 fm) in $^6{\rm Li}$ to 2.25
fm (2.29 fm ) in $^7_\Lambda{\rm Li}$. We also see from the table
that the change of $r_n$ and $r_p$ gradually decreases with
increasing mass number. The above RMF results reveal the
universality of the shrinkage effect for $\Lambda$-hypernuclei,
but not for $\Xi$-hypernuclei. It is particular interesting to
observe a quite different effect caused by $\Xi$ hyperon impurity.

\section{The effect of the $\Xi$ impurity}

In order to see whether or not there is the shrinkage effect of
$\Xi$-hypernuclei, we have carried out the standard RMF
calculations for some $\Xi^-$- and $\Xi^0$-hypernuclei. Due to
insufficient experiment information on $\Xi$-hypernuclei, the
$\Xi$ potential well depth is relatively uncertain, values
appearing in the literature range from about -30 to -10 MeV.
Recent experiments with light nuclei suggest that the value lies
on the less bound size of this range $^{\cite{6025,6024}}$.
However, it may be more deeply bound for heavy nuclei
$^{\cite{new20}}$. As a result, a number of values of the $\Xi$
potential well depth $V^\Xi_0$ for each hypernuclei are used to
test the sensitivity of the position of $\Xi$ single-particle
energy levels to the potential depth. In Fig.2, we only present
results of calculations for the nucleus Zr with a comparison of
the $\Xi^-$ (upper part) and the $\Xi^0$ (lower part)
single-particle levels. It can be seen that the change of the
potential well depth causes large change in the single-particle
energies. As $V^\Xi_0$ deeper, the single-particle energies of
hyperon increase significantly, and the spin-orbit splitting has a
little larger. We also see that the attractive Coulomb interaction
for $\Xi^-$ leads to a considerable stronger binding of $\Xi^-$ in
nuclei when compared with $ \Xi^0$-hypernuclei. In Fig.3, we give
the $\Xi^-$(upper part) and $\Xi^0$(lower part) binding energies
in the nuclei O, Ca, Zr, Pb for $V^\Xi_0=-10$ and $-28$ MeV (only
the $1s, 1p, 1d, 1f$ states are given). The solid (dashed) curves
are the results for $V^\Xi_0=-10(-28)$MeV.

Next let us go further into the question how the static properties
of the $\Xi$-hypernuclei are affected by the potential depth
$V^\Xi_0$. Both $\Xi^-$ and $\Xi^0$ are at the 1$s$ state in
hypernuclei. Our RMF results are shown in table \ref{t2} with
$V^\Xi_0=$-10, -18 and -28 MeV, respectively.
 As seen from the table \ref{t2},
with increasing depth $|V^\Xi_0|$ from 10 MeV to 28 MeV, the
binding energies per baryon ($-E/A$) become larger. We can see
such an uncertainty of $\Xi$ potential well depth gets clearly
reflected in an important variation of the $\Xi$ binding energies,
as it is shown in Figs. 2 and 3 and also of the binding energies
per baryon ($-E/A$), presented in Table \ref{t2}. Because of that,
no firm conclusions can be drawn from the quoted values of $-E/A$.
We can also notice that the charge radius and the r.m.s. radii of
the $\Xi$ hyperon, neutrons and protons become smaller with
increasing potential well depth. Note that the reduction of r.m.s.
radius for the neutrons ($r_n$) and protons ($r_p$) is different.
In the case of $\Xi^-$-hypernuclei, the reduction of the $r_p$ is
faster than that of the $r_n$. While in the case of
$\Xi^0$-hypernuclei, the reduction of the $r_p$ is slower that
that of the $r_n$. Thus the RMF model predicts that the proton and
nutron distribution have different response to the potential depth
$V_0^\Xi$ for the $\Xi^-$- and $\Xi^0$-hypernuclei.

Now, we study whether the $\Xi$ hyperon impurity has the gluelike
role as the $\Lambda$ does. The results are shown in table
\ref{t1} in the form $C^{+A}_{+B}$, where the central values ($C$)
are the results obtained with the -18 MeV $\Xi$ potential well
depth, while the extremes of the uncertainty interval $C+A$ and
$C+B$ are obtained with $V^\Xi_0=$-10 MeV and -28 MeV,
respectively. A similar presentation is used in table \ref{t3}
(where rho exchange is not considered, i.e., $g_{\rho \Xi}=0$).
 From table \ref{t1}, we find, by adding a $\Xi^-$ hyperon to the nuclei, the r.m.s.
radius of the neutrons become a little larger, while r.m.s. radius
of the protons become much smaller, comparing with that in the
normal nuclei. Contrast to the situation of $\Xi^-$-hypernuclei,
by adding a $\Xi^0$ hyperon, the r.m.s. radius of the protons
become larger and that of the neutrons become smaller. This is
different from the situation of adding a $\Lambda$ hyperon. We
know that $\Lambda$ has a gluelike role, both the r.m.s. radii of
the protons and neutrons become smaller when adding a $\Lambda$.
Note that $\Lambda$, $\Xi^-$ and $\Xi^0$ are different particles
from proton and neutron, they are all not constrained by the Pauli
exclusion, it is obviously that the common explanation for the
shrinkage does not suit the case of $\Xi^-$ and $\Xi^0$.
Otherwise, both $\Lambda$ and $\Xi^0$ hyperon are neutral, hence
the origin of the above difference can not be attributed to the
Coulomb potential. There must be some other source that we don't
recognized.

To reach a better understanding of the different behavior of the
$\Lambda, \Xi^-$ and $\Xi^0$ impurities in the nucleus, we make an
inspection of their isospin. $\Lambda$, $\Xi^-$ and $\Xi^0$ have
different third component of isospin, which may be responsible for
their different behavior. The different third component of isospin
works through the coupling of baryon with the $\rho$ mesons in the
RMF model. We may imagine if the $\rho$ mesons couplings for
$\Xi^-$ and $\Xi^0$ are omitted from the RMF calculation
($g_{\rho\Xi}=0$), the above mentioned different behavior of
$\Xi^-$ and $\Xi^0$ shall disappear. After eliminating the
contribution of the $\rho$ mesons, the RMF results are shown in
table \ref{t3}, from which we find the r.m.s. radii of both the
protons and neutrons reduce when adding a $\Xi^-$ or $\Xi^0$
hyperon to the normal nuclei, the same as the situation of adding
a $\Lambda$ hyperon. We obtain the same nuclear shrinkage by
$\Xi^-$ and $\Xi^0$ when ignoring the contribution of the $\rho$
mesons. From the interactive term of nucleons with the $\rho$
mesons, we can find : when adding a $\Xi^-$, the attractive force
increases for the protons and the repulsive force increase for the
neutrons, the situation is contrary to the above when adding a
$\Xi^0$. That explains the above RMF results reasonably. So, we
can conclude the $\rho$ mesons play an important role, and the
different behavior of the $\Lambda$, $\Xi^-$ and $\Xi^0$
impurities is due to their different isospin. Although the changes
are small, the different response of $r_p$ and $r_n$ to $\Xi^-$
and $\Xi^0$ may be interesting to know what kind of properties the
two-body $\Xi N$ interaction. Probably the isospin $T=0$
interaction is attractively large, while the $T=1$ interaction is
repulsive and small. Although the r.m.s. radius is reduced only
for one kind of nucleons, but the r.m.s radius of other kind of
nucleons become larger, that is seems that the nuclei may even
swell somewhat when adding a $\Xi^-$ or $\Xi^0$. That is very
different from the nuclear shrinkage by $\Lambda$.

\section{Summary and conclusion}

Within the framework of the RMF theory, the $\Lambda$
single-particle energies was calculated and the results are in
good agreement with the experiments for all of the hypernuclei
considered. Very small spin-orbit splitting for
$\Lambda$-hypernuclei are observed, which is agreement with
earlier phenomenological analysis. From the investigation of the
effects of $\Lambda$ on the core nucleus, We obtain the shrinkage
effect inducing by $\Lambda$ hyperon impurity, otherwise, we find
other light and medium $\Lambda$-hypernuclei also have this
shrinkage effect, i.e., the gluelike role of $\Lambda$ impurity is
universal.

 For $\Xi$-hypernuclei, first, we study the effect of the potential well
depth $V^\Xi_0$ on the static properties of $\Xi$-hypernuclei. We
can see the uncertainty of $\Xi$ potential well depth gets clearly
reflected in an important variation of the $\Xi$ binding energies,
because of that, no firm conclusions can be drawn from the quoted
$\Xi$ binding energies and values of $-E/A$. In the
$\Xi^-$-hypernuclei, the reduction of r.m.s. radius of the protons
is larger than the reduction of that of the neutrons, while in the
$\Xi^0$-hypernuclei, the reduction of r.m.s. radius of the
neutrons are larger than that of the protons with the deeper
potential well depth. The strength of the effect of $V^\Xi_0$ on
different nucleons is different in $\Xi$-hypernuclei.
 The effect of $V^\Xi_0$ on the hypernuclei decreases with
the increasing of the atomic number.

After that, we study the effect of the adding $\Xi$ hyperon on the
nuclear core, we find: by adding a $\Xi^-$ hyperon to the nucleus,
the r.m.s. radius of the neutrons become a little larger, while
the r.m.s. radius of the protons become smaller, comparing with
that in the normal nucleus, and the decrease of the r.m.s. radius
of the protons is larger as the $V^\Xi_0$ deeper. While when
adding a $\Xi^0$ hyperon, the r.m.s. radius of the protons become
a little larger and that of the neutrons become smaller. Although
the r.m.s. radius is reduced only for one kind of nucleons, but
the r.m.s radius of other kind of nucleons become larger, that is
seems that the nuclei may even swell somewhat when adding a
$\Xi^-$ or $\Xi^0$. That is very different from the nuclear
shrinkage by $\Lambda$. And we find the $\rho$ mesons play an
important role, the different effect on the nuclear core by
$\Lambda$, $\Xi^-$, $\Xi^0$ is due to their different isospin.
Although the changes are small, the different response of $r_p$
and $r_n$ to $\Xi^-$ and $\Xi^0$ may be interesting to know what
kind of properties of the two-body $\Xi N$ interaction. Probably
the isospin $T=0$ interaction is attractively large, while the
$T=1$ interaction is repulsive and small.

The present work only focuses on the pure $\Lambda$ and $\Xi$
hypernuclei, the coupling between $\Xi N$ and $\Lambda \Lambda$
channels in $\Xi$ hypernuclei isn't taken into consideration. The
physics of $\Lambda \Lambda$ hypernuclei ($\Lambda \Lambda$ and
$\Xi N$ mix up in a formalism of coupled channel) and $\Lambda
\Xi$ hypernuclei have attracted a lot of attention$^{\cite{ll}}$
and are subject of current investigation, because of that, more
reliable information on $\Xi N$ interaction and $\Xi$-nucleus are
desired.

 \hspace{5cm}{\Large
\bf Acknowledgments}

 This work was supported in part by China
postdoctoral science foundation (2002032169), National Natural
Science Foundation of China (10275037) and China Doctoral
 Programme Foundation of Institution of Higher Education
 (20010055012.).
We would like also to thank Prof. Chonghai Cai, Prof. Lei Lee  and
Baoxi Sun for useful discussions.
\medskip

\newpage

\begin{figure}[pthb]
\centering\includegraphics[width=8.cm,height=6cm]{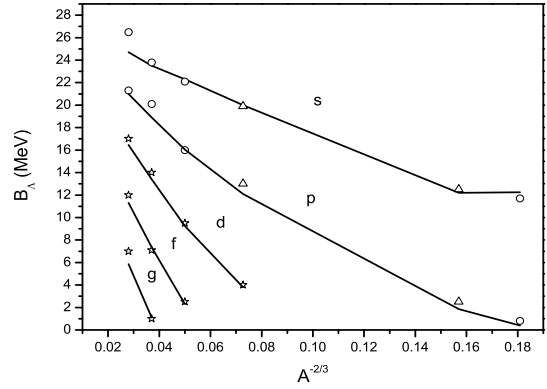}
\caption{ $\Lambda$ binding energies (in MeV)
 for some hypernuclei($^{13}_\Lambda \rm C$, $^{16}_\Lambda \rm O$, $^{51}_\Lambda \rm V$, $^{89}_\Lambda \rm Y$ , $^{139}_\Lambda \rm La$, $^{208}_\Lambda  \rm Pb
 $).
The solid lines are our RMF results with parameters of
Ref.\cite{ma}.
 The experimental
 data are taken from Ref.\cite{7}, Ref.\cite{1999},
 and Ref.\cite{pile} denoted by $\triangle, \circ, \star$, respectively. }\label{f1}
\end{figure}

\begin{figure}[pthb]
\centering\includegraphics[width=8.cm,height=6cm]{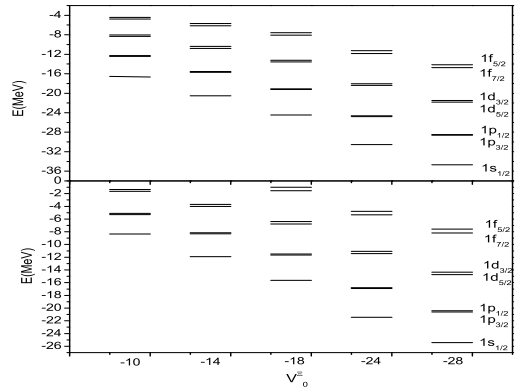}
\caption{  Dependence of the position of the $\Xi^-$ (upper part)
and the $\Xi^0$ (lower part) single-particle levels in Zr for
$V^\Xi_0=-10, -14, -18, -24, -28$ MeV (only the $1s, 1p, 1d,1f$
states are given). }\label{f2}
\end{figure}

\begin{figure}[pthb]
\centering\includegraphics[width=8.cm,height=6cm]{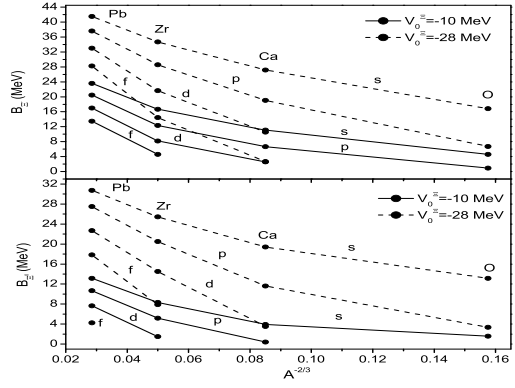}
\caption{ Comparison of the $\Xi^-$ (upper part) and the $\Xi^0$
(lower part) binding energies in O, Ca, Zr, Pb for $V^\Xi_0=-10$
and -28 MeV.}\label{f3}
\end{figure}

\newpage

\begin{table}
 \caption{Binding energy per baryon, -$E/A$ (in MeV),
 r.m.s. charge radius $r_{ch}$(those of the nucleons, in fm),
r.m.s. radii of the hyperon ($\Lambda, \Xi^-, \Xi^0$), neutron
 and proton, $r_{\rm y}$, $r_n$ and $r_p$
 (in fm), respectively, including the contribution of the $\rho$ mesons.
   The configuration
of hyperon is $1s_{1/2}$ for all hypernclei. The meaning of $Z$ in
$^A Z$ is the number of protons. The results of $\Xi$-hypernuclei
are given in the form $C^{+A}_{+B}$, where $C$, $C+A$ and $C+B$
are the results obtained with $V^\Xi_0=$-18, -10 and -28 MeV,
respectively.} \label{t1}
 {\small\begin{center}
 \begin{tabular}{cccccc|ccccccc}  \hline
                      $^AZ$&$-E/A$ &$r_{ch}$&$r_{\rm y}$&$r_n$&$r_p$
                      &     $^AZ$            &$-E/A$ &$r_{ch}$&$r_{\rm y}$&$r_n$&$r_p$ \\
                      \hline
                      $^6{\rm Li}$&5.67&2.52&&2.32&2.37 &$^{16}{\rm O}$                     &  8.04 &2.70    & &2.55
 &2.58\\
     $^7_\Lambda{\rm Li}$&5.47&2.43&2.58&2.25&2.29 &  $^{17}_\Lambda{\rm O}$ &8.27&2.70&2.43&2.55&2.58\\
$^7_{\Xi^-}{\rm Li}$&$5.18^{-0.27}_{+0.58}$&$2.39^{+0.07}_{-0.08}$&$3.20^{+1.55}_{-0.90}$&$2.35^{+0.00}_{-0.04}$&$2.25^{+0.07}_{-0.10}$&$^{17}_{\Xi^-}{\rm O}$                               &$8.14^{-0.29}_{+0.45}$  &$2.67^{+0.02}_{-0.02}$    &$2.59^{+0.74}_{-0.43}$       &$2.58^{-0.01}_{-0.01}$ &$2.55^{+0.01}_{-0.03}$  \\
 $^7_{\Xi^0}{\rm Li}$& $4.99^{-0.22}_{+0.56}$&$2.55^{-0.01}_{-0.04}$&$3.49^{+2.37}_{-1.13}$&$2.23^{+0.06}_{-0.09}$&$2.41^{-0.01}_{-0.04}$&                     $^{17}_{\Xi^0}{\rm O}$          &$7.92^{-0.26}_{+0.44}$    &$2.73^{-0.01}_{+0.00}$
 &$2.71^{+1.14}_{-0.51}$
 &$2.53^{+0.01}_{-0.03}$
 &$2.60^{+0.00}_{-0.00}$
 \\\hline
$^8{\rm Be}$&5.42&2.48&&2.30&2.34&$^{40}{\rm Ca}$& 8.52  &3.46 &
&3.31 &3.36      \\
      $^9_\Lambda{\rm Be}$&5.58&2.44&2.40&2.27&2.30 &  $^{41}_\Lambda{\rm Ca}$ &8.75&3.46&2.70&3.31&3.36    \\
  $^9_{\Xi^-}{\rm Be}$ &$5.30^{-0.32}_{+0.62}$&$2.41^{+0.04}_{-0.07}$&$2.80^{+1.22}_{-0.68}$&$2.33^{+0.00}_{-0.03}$&$2.26^{+0.04}_{-0.06}$&                    $^{41}_{\Xi^-}{\rm Ca}$          &$8.75^{-0.16}_{+0.23}$     &$3.44^{+0.00}_{-0.02}$   &$2.75^{+0.49}_{-0.30}$       &$3.33^{+0.00}_{-0.01}$   &$3.34^{+0.01}_{-0.02}$     \\
    $^9_{\Xi^0}{\rm Be}$ &$5.09^{-0.27}_{+0.60}$&$2.50^{+0.00}_{-0.02}$&$2.99^{+1.91}_{-0.83}$&$2.25^{+0.04}_{-0.07}$&$2.36^{+0.00}_{-0.02}$&                      $^{41}_{\Xi^0}{\rm Ca}$           &$8.56^{-0.15}_{+0.23}$  &$3.47^{+0.00}_{-0.00}$  &$2.87^{+0.72}_{-0.37}$
    &$3.30^{+0.00}_{-0.02}$
    &$3.37^{+0.00}_{-0.00}$
    \\\hline$^{12}{\rm C}$                  &7.47    &2.46         && 2.30
&2.32  &$^{208}{\rm Pb}$&              7.90  &5.51  & &5.71 &5.45
\\
 $^{13}_\Lambda{\rm C}$ &7.79&2.44&2.19&2.28&2.30&   $^{209}_\Lambda{\rm Pb}$ &7.98&5.51&4.05&5.71&5.44    \\
$^{13}_{\Xi^-}{\rm C}$          &$7.55^{-0.37}_{+0.65}$
&$2.42^{+0.02}_{-0.04}$ &$2.43^{+0.93}_{-0.51}$
&$2.32^{+0.00}_{-0.02}$ &$2.27^{+0.03}_{-0.04}$  &                    $^{209}_{\Xi^-}{\rm Pb}$          &$8.01^{-0.03}_{+0.05}$     &$5.50^{+0.00}_{-0.01}$   &$3.68^{+0.19}_{-0.13}$       &$5.72^{+0.00}_{-0.01}$   &$5.44^{+0.00}_{-0.01}$     \\
$^{13}_{\Xi^0}{\rm C}$          &$7.31^{-0.33}_{+0.63}$
&$2.48^{+0.00}_{-0.01}$ &$2.56^{+1.46}_{-0.60}$
&$2.26^{+0.03}_{-0.04}$ &$2.34^{+0.00}_{-0.01}$ &
$^{209}_{\Xi^0}{\rm Pb}$ &$7.96^{-0.04}_{+0.05}$
&$5.51^{+0.00}_{-0.00}$ &$4.05^{+0.29}_{-0.20}$
&$5.70^{+0.01}_{-0.00}$ &$5.45^{+0.00}_{-0.00}$
                      \\\hline
\end{tabular}
\end{center}}
\end{table}

\newpage

\begin{table}
\caption{Binding energy per baryon, -$E/A$ (in MeV), r.m.s. charge
radius $r_{ch}$(those of the nucleons, in fm),
 r.m.s. radii of the hyperon, neutron
 and proton, $r_y$, $r_n$ and $r_p$ (in fm), respectively, including the contribution of the $\rho$ mesons.
The meaning of $Z$ in $^A Z$ is the number of protons in
hypernuclei. The configuration of hyperon is $1s_{1/2}$ for all
hypernuclei. The results of $\Xi$ hypernuclei are given for
$V^\Xi_0=-10, -18, -28$ MeV.} \label{t2}
 {\small\begin{center}
 \begin{tabular}{ccccccc|cccccccc}  \hline
 $^AZ$ & $V^\Xi_0$&$-E/A$ &$r_{ch}$&$r_{\rm y}$&$r_n$&$r_p$ &$^A Z$& $V_0^\Xi$&$-E/A$ &$r_{ch}$&$r_{\rm y}$&$r_n$&$r_p$ \\ \hline
   $^6{\rm Li}$&&5.67&2.52&&2.32&2.37 &$^{16}{\rm O}$&                     &  8.04 &2.70    & &2.55
   &2.58\\\hline
$^7_{\Xi^-}{\rm Li}$&-10&4.91&2.46&4.75&2.35&2.32&$^{17}_{\Xi^-}{\rm O}$&-10 &7.85   &2.69    &3.33       &2.57 &2.56 \\
&-18&5.18&2.39&3.20&2.35&2.25&  &-18 &8.14  &2.67&2.59       &2.58 &2.55 \\
 &-28&5.76&2.31&2.30&2.31&2.15&&-28   &8.59    &2.65    &2.16       &2.57 &2.52  \\\hline
 $^7_{\Xi^0}{\rm Li}$&-10& 4.77&2.54&5.86&2.29&2.40&$^{17}_{\Xi^0}{\rm O}$&-10&7.66   &2.72   &3.85       & 2.54  &2.60     \\
&-18& 4.99&2.55&3.49&2.23&2.41&     &-18          &7.92    &2.73   &2.71       &2.53   &2.60     \\
&-28&5.55&2.51&2.36&2.14&2.37&&-28&8.36&2.73&2.20&2.50&2.60\\\hline\hline
 $^8{\rm Be}$&&5.42&2.48&&2.30&2.34&$^{40}{\rm Ca}$&& 8.52 &3.46 &&3.31 &3.36
 \\\hline
  $^9_{\Xi^-}{\rm Be}$&-10&4.98&2.45&4.02&2.33&2.30&$^{41}_{\Xi^-}{\rm Ca}$&-10           &8.59   &3.44   &3.24       &3.33   &3.35     \\
  &-18&5.30&2.41&2.80&2.33&2.26&                    &-18          &8.75     &3.44   &2.75       &3.33   &3.34     \\
&-28&5.92&2.34&2.12&2.30&2.20 & &-28&8.98&3.42   &2.45 &3.32 &3.32
\\  \hline
    $^9_{\Xi^0}{\rm Be}$&-10&4.82&2.50&4.90&2.29&2.36&$^{41}_{\Xi^0}{\rm Ca}$&-10          &8.41  &3.47  &3.59        &3.30   &3.37    \\
    &-18&5.09&2.50&2.99&2.25&2.36&                      &-18           &8.56  &3.47  &2.87        &3.30   &3.37     \\
    &-28&5.69&2.48&2.16&2.18&2.34&                      &-28           &8.79  &3.47  &2.50        &3.28   &3.37     \\
                      \hline
   \hline$^{12}{\rm C}$  &                &7.47    &2.46         && 2.30
&2.32  &$^{208}{\rm Pb}$& &             7.90  &5.51  & &5.71 &5.45
\\\hline
$^{13}_{\Xi^-}{\rm C}$&-10        &7.18   &2.44    &3.36 &2.32
&2.30 &$^{209}_{\Xi^-}{\rm Pb}$&-10            &7.98   &5.50   &3.87       &5.72   &5.44     \\
                      &-18         &7.55  &2.42    &2.43       &2.32 &2.27
                      &                    &-18          &8.01     &5.50   &3.68       &5.72   &5.44     \\
                     &-28         &8.20    &2.38    &1.92       &2.30 &2.23
                     &                    &-28          &8.06     &5.49   &3.55       &5.71   &5.43     \\  \hline
$^{13}_{\Xi^0}{\rm C}$&-10         &6.98   &2.48   &4.02       &
2.29  &2.34              &$^{209}_{\Xi^0}{\rm Pb}$&-10         &7.92  &5.51  &4.34        &5.71   &5.45    \\
                     &-18          &7.31    &2.48   &2.56       &2.26   &2.34     &                      &-18           &7.96  &5.51  &4.05        &5.70   &5.45     \\
                      &-28         &7.94   &2.47   &1.96       &2.22   &2.33  &                      &-28           &8.01 &5.51  &3.85        &5.70   &5.45        \\
                      \hline\hline

\end{tabular}
 \end{center}}
\end{table}

\newpage

\begin{table}[pthb]
 \caption{Binding energy per baryon, -$E/A$ (in MeV),
 r.m.s. charge radius $r_{ch}$(those of the nucleons, in fm),
 r.m.s. radii of the hyperon, neutron
 and proton, $r_y$, $r_n$ and $r_p$
 (in fm), respectively, without the contribution of the $\rho$ mesons.
The meaning of $Z$ in $^A Z$ is the number of protons.   The
configuration of hyperon is $1s_{1/2}$ for all hypernuclei. The
results of $\Xi$-hypernuclei are given in the form $C^{+A}_{+B}$,
where $C$, $C+A$ and $C+B$ are the results obtained with
$V^\Xi_0=$-18, -10 and -28 MeV, respectively.} \label{t3}
 {\small\begin{center}
 \begin{tabular}{cccccc|ccccccc}  \hline
$^A Z$&$-E/A$ &$r_{ch}$&$r_{\rm y}$&$r_n$&$r_p$ &          $^A Z$&
$-E/A$ &$r_{ch}$&$r_{\rm y}$&$r_n$&$r_p$ \\
\hline
 $^6{\rm Li}$&5.67&2.52&&2.32&2.37 &$^{16}{\rm O}$        &               8.04 &2.70    & &2.55 &2.58
\\
$^7_{\Xi^-}{\rm
Li}$&$5.27^{-0.32}_{+0.68}$&$2.44^{+0.05}_{-0.07}$&$2.82^{+1.35}_{-0.75}$&$2.27^{+0.04}_{-0.07}$&$2.30^{+0.05}_{-0.08}$&$^{17}_{\Xi^-}{\rm O}$   &$8.19^{-0.31}_{+0.47}$  &$2.70^{+0.01}_{-0.02}$    &$2.46^{+0.67}_{-0.39}$       &$2.55^{+0.01}_{-0.02}$ &$2.57^{+0.01}_{-0.01}$  \\
 $^7_{\Xi^0}{\rm Li}$& $5.06^{-0.27}_{+0.64}$&$2.46^{+0.04}_{-0.07}$&$3.04^{+2.12}_{-0.91}$&$2.28^{+0.03}_{-0.07}$&$2.32^{+0.05}_{-0.08}$&$^{17}_{\Xi^0}{\rm O}$    &$7.94^{-0.27}_{+0.46}$    &$2.70^{+0.01}_{-0.01}$   &$2.60^{+1.03}_{-0.48}$       &$2.55^{+0.01}_{-0.01}$   &$2.58^{+0.00}_{-0.02}$     \\\hline
$^8{\rm Be}$&5.42&2.48&&2.30&2.34&$^{40}{\rm Ca}$& 8.52  &3.46 &
&3.31 &3.36      \\
  $^9_{\Xi^-}{\rm Be}$  &$5.39^{-0.37}_{+0.69}$&$2.44^{+0.03}_{-0.05}$&$2.54^{+1.05}_{-0.59}$&$2.28^{+0.02}_{-0.05}$&$2.30^{+0.03}_{-0.05}$&$^{41}_{\Xi^-}{\rm Ca}$           &$8.77^{-0.17}_{+0.24}$     &$3.45^{+0.01}_{-0.01}$   &$2.65^{+0.45}_{-0.29}$       &$3.31^{+0.00}_{-0.01}$   &$3.35^{+0.01}_{-0.01}$     \\
    $^9_{\Xi^0}{\rm Be}$ &$5.16^{-0.33}_{+0.66}$&$2.45^{+0.03}_{-0.05}$&$2.71^{+1.62}_{-0.71}$&$2.28^{+0.03}_{-0.05}$&$2.31^{+0.03}_{-0.05}$&$^{41}_{\Xi^0}{\rm Ca}$  &$8.56^{-0.15}_{+0.23}$  &$3.46^{+0.00}_{-0.01}$  &$2.84^{+0.74}_{-0.37}$        &$3.31^{+0.00}_{-0.01}$  &$3.36^{+0.00}_{-0.01}$    \\\hline
$^{12}{\rm C}$                 &7.47&2.46&&2.30&2.32 &$^{208}{\rm
Pb}$          &7.90  &5.51  & &5.71&5.45      \\
 $^{13}_{\Xi^-}{\rm C}$         &$7.63^{-0.41}_{+0.71}$  &$2.44^{+0.02}_{-0.03}$    &$2.26^{+0.82}_{-0.47}$
 &$2.28^{+0.01}_{-0.03}$
 &$2.30^{+0.02}_{-0.03}$
 &$^{209}_{\Xi^-}{\rm Pb}$                             &$8.00^{-0.07}_{+0.09}$     &$5.50^{+0.01}_{-0.00}$   &$4.64^{+2.15}_{-1.22}$       &$5.71^{+0.00}_{-0.01}$   &$5.45^{+0.00}_{-0.01}$     \\
$^{13}_{\Xi^0}{\rm C}$   &$7.37^{-0.37}_{+0.68}$
&$2.45^{+0.01}_{-0.03}$ &$2.38^{+1.24}_{-0.55}$
&$2.29^{+0.01}_{-0.04}$
&$2.31^{+0.01}_{-0.03}$&$^{209}_{\Xi^0}{\rm Pb}$
&$7.93^{-0.03}_{+0.05}$ &$5.51^{+0.00}_{-0.01}$
&$4.17^{+0.47}_{-0.26}$ &$5.71^{+0.00}_{-0.00}$
&$5.44^{+0.01}_{-0.00}$
\\\hline

\end{tabular}
\end{center}}
\end{table}
\end{document}